\documentclass[aps,pra,twocolumn,showpacs,floatfix, superscriptaddress,longbibliography,noeprint]{revtex4-2}
\usepackage[utf8]{inputenc} 

\usepackage{amssymb,amsmath}
\usepackage{graphicx}
\usepackage{xcolor}
\usepackage{dsfont}
\usepackage{hyperref}
\hypersetup{
	colorlinks=true,       
	linkcolor=blue,          
	citecolor=green,       
	filecolor=magenta,      
	urlcolor=blue          
}
\usepackage[caption=false]{subfig}

\def\Rb87{^{87}\mathrm{Rb}}                     
\def\K40{^{40}\mathrm{K}}                    		

\def\ex{\mathbf{e}_x}  
  
\def\ez{\mathbf{e}_z}  

\def\ket#1{\mathinner{|{#1}\rangle}}

\begin{document}

\title{Feedback-stabilized dynamical steady states in the Bose-Hubbard model}


\author{Jeremy T. Young}
\email{jeremy.young@colorado.edu}
\affiliation{JILA, NIST and Department of Physics, University of Colorado, Boulder, Colorado 80309, USA}
\affiliation{Center for Theory of Quantum Matter, University of Colorado, Boulder, Colorado 80309, USA}

\author{Alexey V. Gorshkov}
\affiliation{Joint Quantum Institute, NIST/University of Maryland, College Park, Maryland 20742, USA}
\affiliation{Joint Center for Quantum Information and Computer Science, NIST/University of Maryland, College Park, Maryland 20742, USA}

\author{I. B. Spielman}
\email{ian.spielman@nist.gov}
\affiliation{Joint Quantum Institute, NIST/University of Maryland, College Park, Maryland 20742, USA}
\homepage{http://ultracold.jqi.umd.edu}
\date{\today}

\begin{abstract}
The implementation of a combination of continuous weak measurement and classical feedback provides a powerful tool for controlling the evolution of quantum systems. In this work, we investigate the potential of this approach from three perspectives. First, we consider a double-well system in the classical large-atom-number limit, deriving the exact equations of motion in the presence of feedback. Second, we consider the same system in the limit of small atom number, revealing the effect that quantum fluctuations have on the feedback scheme. Finally, we explore the behavior of modest sized Hubbard chains using exact numerics, demonstrating the near-deterministic preparation of number states, a tradeoff between local and non-local feedback for state preparation, and evidence of a feedback-driven symmetry-breaking phase transition.
\end{abstract}

\maketitle

Equilibrium is a concept central to many-body physics, both classical and quantum.  It is not surprising then, that understanding new kinds of equilibrium and discovering breakdowns to equilibration are driving significant progress in physics.  Recent examples include the generalized Gibbs ensemble describing the limited equilibration possible for  integrable quantum systems~\cite{Cassidy2011,Langen2015,Ilievski2016,Calabrese2016,Vidmar2016,Mori2018,Kinoshita2006,Gring2012,Tang2018}, many-body states in driven-dissipative quantum systems~\cite{Diehl2008,Szymanska2007,Yi2012,Bardyn2013, Baumann2010,Carr2013a,Malossi2014,Jin2016,Rodriguez2017, Fitzpatrick2017,Dogra2019,Scarlatella2021}, while quantum glasses exhibit extremely long relaxation times \cite{Wu1991,Gopalakrishnan2011} and many-body localized systems never equilibrate \cite{Altman2015b,Nandkishore2015,Gogolin2016,Alet2018,Abanin2019, Schreiber2015,Choi2016,Smith2016,Roushan2017,Guo2020a}.  Here we focus instead on many-body quantum systems maintained in dynamical steady state---a kind of generalized equilibrium---stabilized by the interplay of unitary dynamics, minimally destructive quantum measurement, and classical feedback \cite{Lloyd2000,Thomsen2002,Shankar2019,Muschik2013, MunozArias2020a,Wu2021,Hurst2019,Hurst2020, Grimsmo2014,Kopylov2015,Mazzucchi2016, MunozArias2020b,Ivanov2020,Kroeger2020,Ivanov2020b,Ivanov2021}.

Statistical mechanical equilibrium is based on the straightforward observation that a system's eigenstates are probabilistically occupied according to the thermal Boltzmann distribution, subject to any relevant constraints.  
The resulting density operator affords a time-independent description of the system as a thermal-equilibrium steady state.  
Conversely, any system described by a time-independent density operator may be associated with thermal equilibrium, for an effective Hamiltonian proportional to the logarithm of that density operator, although detailed balance may not necessarily be satisfied.
In this sense then, is it possible to identify and explore systems maintained in low-entropy steady states associated with exotic effective Hamiltonians with highly non-local or $N$-body interactions? Optical pumping~\cite{Happer1972} and laser cooling~\cite{Lett1988}---both described by the physics of open quantum systems---are iconic examples where large ensembles of atoms enter low-entropy steady states well described by single-atom physics with little correlation between atoms.

In contrast, quantum error correction (QEC) codes are highly specialized and sophisticated examples of measurement-feedback systems that can dynamically stabilize strongly entangled arrays of qubits \cite{Shor1995,Terhal2015}.  A digital quantum computation device consists of an interconnected collection of qubits, and QEC utilizes multi-qubit syndrome measurements, often realized via ``ancilla'' qubits that are first coupled to the error-prone qubits and then measured.
The coupling and measurement are designed to be sensitive to select errors, but not to the quantum state involved in computation.
In some forms of QEC~\cite{Devitt2013}, the classical information resulting from measuring the ancilla  can inform subsequent error-correcting feedback stages, thereby maintaining the quantum computation device in a type of dynamical steady state.

Quantum state preparation is closely related to QEC: both attempt to drive a system towards a particular state. In the case of QEC, this is the state prior to an error; in the case of state preparation, it is a particular state of interest. State preparation is particularly important in the field of metrology \cite{Giovannetti2011,Degen2017}. By preparing highly entangled states, such as squeezed states \cite{Walls1983a,Kitagawa1993,Thomsen2002,SchleierSmith2010,Hosten2016,Cox2016,Shankar2019}, it becomes possible to make measurements that are far more accurate than in an unentangled classical system. While a variety of sophisticated techniques have resulted in experimental measurements with incredible accuracy, incoherent effects often provide a fundamental limitation. Like with typical QEC for quantum computation, measurement and feedback may provide a means of reducing the problems introduced by incoherent processes \cite{Korotkov2001, Ahn2002, Sarovar2004,Dur2014,Kessler2014,Arrad2014,Unden2016,Sekatski2017,Reiter2017,Zhou2018,Layden2019}.

Here, we study this problem from three perspectives progressively increasing in complexity.  
(I) we begin by considering a minimal two-site system in the large-atom-number classical limit---a type of non-linear top---and derive the exact equations of motion including feedback.  
This model predicts what types of feedback can drive the system into different fixed points, providing guidance for what type of quantum states will appear.  
(II) We then consider a stochastic Schr\"odinger equation description of the same double well. By investigating both the classical large-atom limit and the quantum small-atom limit, we connect the classical fixed points to their associated quantum dynamical steady states. 
(III) We conclude with the numerically exact time dynamics for modest sized Hubbard chains in the quantum low-density limit (see Refs.~\cite{Hurst2019} and \cite{Hurst2020} for investigations in the more classical high-density limit) and show that nearly arbitrary distributions of number states can be near-deterministically prepared, and that non-local feedback algorithms can both improve fidelity and accelerate state preparation. 
Moreover, we show that the corresponding feedback can be modified in a simple way to realize a symmetry breaking transition. 

From a broader perspective, this manuscripts shows that straightforward feedback derived from highly idealized continuous quantum measurements can stabilize low-entropy dynamical steady states.  It is the task of future work to expand upon the impact of increasingly realistic experimental parameters.

\section{Model}

\begin{figure}[t]
\includegraphics{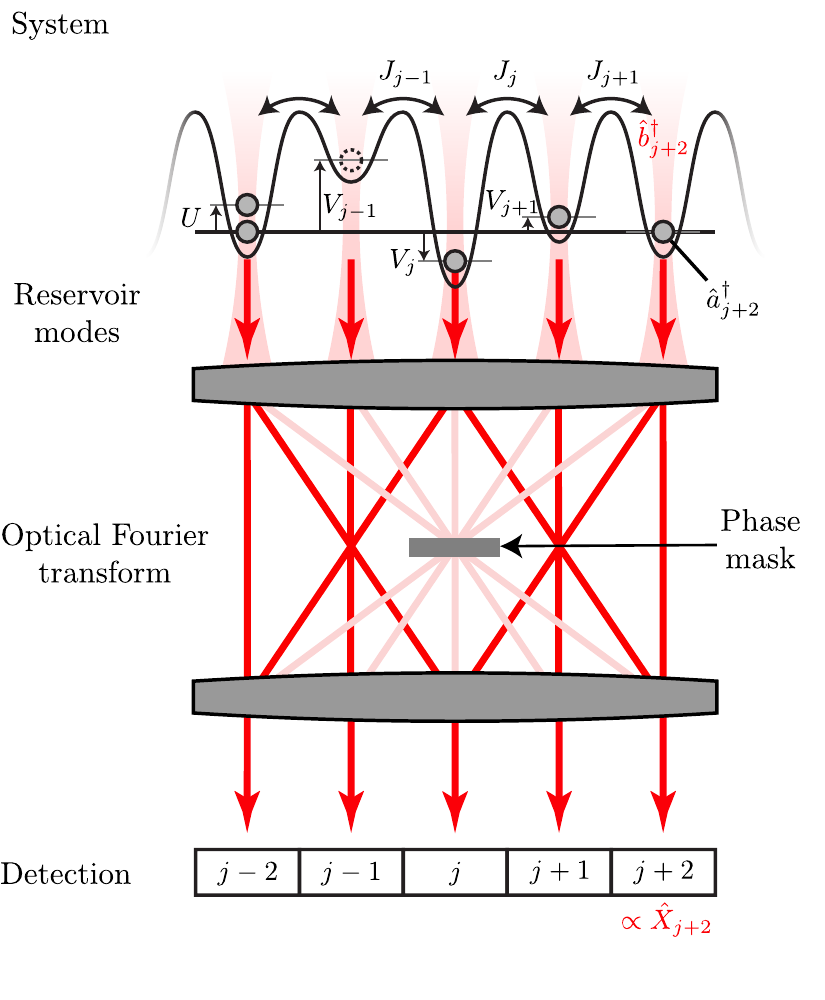}
\caption{System and measurement schematic.  An ensemble of locally interacting (with strength $U$) neutral atoms described by the bosonic field operators $\hat a_j$ are confined in a lattice potential with tunneling strength $J_{j}$ between sites $j$ and $j+1$.  These atoms interact with independent modes of an optical field described by bosonic field operators $\hat b_j$.  After interacting, the optical modes are imaged via phase contrast imaging, implementing a spatially resolved optical homodyne detection providing a measurement sensitive to the quadrature field operator $\hat X_j = (\hat b_j + \hat b^\dagger_j)/2$.  This idealized measurement assumes that the incident laser field is only forward scattered, essentially making the paraxial approximation.}
\label{fig:setup}
\end{figure}

Here we focus on the idealized 1D Bose-Hubbard model subjected to continuous weak measurement \cite{Jacobs2006,Clerk2010} to illustrate the possible dynamical steady states in many-body quantum systems.  As depicted in Fig.~\ref{fig:setup} this model consists of: a 1D Bose-Hubbard chain (the system) dispersedly coupled to a transverse laser field (reservoir); measured via phase contrast imaging (optical homodyne detection); and with local tunneling strengths and on-site potentials that are dynamically adjusted based upon the measurement outcome (feedback).  At time $t$, the Bose-Hubbard chain is governed by the system Hamiltonian
\begin{align*}
\hat H_{\mathcal S} &= \sum_j -J_j \left(\hat a^\dagger_{j+1} \hat a_{j} + {\rm H.c.}\right)  + V_j \hat n_j + \frac{U}{2} \hat n_j \left(\hat n_j - 1\right)
\end{align*}
expressed in terms of the bosonic field operators $\hat a^\dagger_j$ describing the creation of an atom at site $j$ and $\hat{n}_j = a_j^\dagger a_j$ the number operator at site $j$; with positive real-valued tunneling strength $J_j(t)$ between sites $j$ and $j+1$; on-site energy $V_j(t)$; and interaction strength $U$.  The chemical potential is absent in this microcanonical ensemble study.  

{\it Measurement model.} The system is coupled to the reservoir by the system-reservoir Hamiltonian
\begin{align*}
\hat H_{\mathcal SR} &= \hbar g \sum_j \hat a^\dagger_{j} \hat a_{j}\otimes \hat b^\dagger_{j} \hat b_{j},
\end{align*}
that describes a dispersive off-resonance interaction of light with two-level atoms, detected by homodyne measurement~\cite{Hush2013} or equivalently phase contrast imaging, shown in Fig.~\ref{fig:setup}.
Here, the bosonic field operator $\hat b^\dagger_j$ describes the addition of a photon into mode $j$ associated with lattice site $j$ (here we focus on an idealized case with mode functions associated one-to-one with lattice sites);  $g$ captures the strength of the system-reservoir coupling; and the reservoir modes $\hat b_j$ are each in the coherent state $\ket{\alpha}$ prior to interacting with the system.  Physically this operator describes rotations in the $\hat X_j$-$\hat P_j$ quadrature plane for each reservoir mode, in proportion to the local number of atoms.  After interacting with the system for a time $t_m$, the reservoir modes are projectively measured by optical homodyne measurement (in practice implemented by phase contrast imaging) giving access to the $\hat X_j = (\hat b^\dagger_j + \hat b_j)/2$ quadrature observables. 

A strong measurement on the reservoir in effect affords a single weak measurement of the system observables $\hat n_j$ with strength governed by $g$, the reservoir field amplitude $\alpha$, and the measurement time $t_m$.  A measurement outcome
\begin{align}
n_j(t) &= \langle \hat n_j(t) \rangle + \delta n_j
\end{align}
at time $t$ contains a contribution from the system observable's instantaneous expectation value $\langle \hat n_j(t) \rangle$, along with a noise contribution $\delta n_j = m_j / \varphi$ from projectively measuring the reservoir coherent state $\ket{\alpha_j}$.  Here $m_j$ is a classical random variable with zero mean $\overline{m_j}= 0$ and variance $\overline{m_j m_{j^\prime}} = \delta_{j,j^\prime}/2$. 
We consider an optical mode function consisting of a traveling pulse of light with extent $c t_m$, where $c$ is the speed of light. 
Converting from the continuum mode functions to this compact function introduces a factor of $1/(c t_m)$ into the definition of the generalized measurement strength parameter.
This leads to the generalized measurement strength parameter $\varphi^2 = g^2 |\alpha|^2 t_m / c$, defining the measurement noise $\overline{\delta n_j^2} = \Delta n^2 = 1/2 \varphi^2$ and the back-action of this measurement on the system as described by the Kraus operator~\cite{Nielsen2010}
\begin{align}
\hat K(n_j) &=  \exp\left[-\sum_j\frac{\varphi^2 \left(\hat n_j - n_j \right)^2}{2}\right]
\end{align}
conditioned on the measurement outcomes $n_j$. Given an initial state $|\psi\rangle$ and a measurement outcome $n_j$, the post-measurement state of the system is given by $\hat{K}(n_j) |\psi\rangle$.
As one might intuitively expect from a simple classical measurement, this gives a distribution of measurement outcomes with a width $\propto1/\varphi$ that decreases linearly with the system-reservoir coupling $g$ but as the square root of the measurement time $t_m$ and intensity of the probe field $|\alpha|^2$.
In the following analysis, we consider the continuous limit of many such weak measurements, taking the time between subsequent measurements to be $t_m$ and applying measurements with a strength associated with a measurement time $t_m$.

{\it Feedback model.} For sufficiently weak measurements, the quantum projection noise $\delta n_j$ present in any individual measurement can greatly exceed the contribution of the instantaneous expectation value $\langle \hat n_j(t) \rangle$.  As a result, we borrow from classical control theory~\cite{Vegte1994}, and derive an error signal $\epsilon(t)$ from a temporal low-pass filter 
\begin{align}
\tau \dot\epsilon_j(t) + \epsilon_j(t) &= n_j(t), \label{eq:error}
\end{align}
applied to the direct measurement outcomes.  This consists of an integrator with a low-frequency gain limit, with time constant $\tau$. This filter retains the low frequency $\lesssim 1/\tau$ components of $n_j(t)$ (containing contributions from both system dynamics and projection noise), but rejects the high frequency components (dominated by projection noise). 
Here we assume that the measurement time $t_m\ll\tau$, implicitly making Eq.~\eqref{eq:error} a stochastic differential equation, see Ref.~\cite{Hurst2020} for a more detailed discussion.
The resulting error signal $\epsilon_j(t)$ thus approximates the continuous measurement limit of a sequence of weak measurements of $n_j(t)$ with measurement times $\tau$ regardless of the physical measurement time $t_m \ll \tau$. Note that because this filter is applied in real time, $\epsilon_j(t)$ will lag behind $n_j(t)$ by $\tau$, and we take $\epsilon_j(0) = 0$ in our simulations. Because the goal is neither to perform quantum state estimation nor to drive the system into a predefined state, we do not employ Kalman filter techniques~\cite{Kalman1960}.

Here we focus on possible dynamical steady states when this classical information is then fed back into the Hamiltonian in one of two ways. We shift either the on-site energy
\begin{align}
V_j(t) &= V_0 + \delta V_j(t) = V_0 + G_V \epsilon_j(t) \label{eq:FeedbackV}
\end{align}
in proportion to the local feedback signal (with gain $G_V$), or the tunneling 
\begin{align}
J_j(t) &= J_0 + \delta J_j(t) =  J_0 + G_J [\epsilon_j(t)\! -\! \epsilon_{j+1}(t)]^2 \label{eq:FeedbackJ}
\end{align}
based on the difference along that link (with gain $G_J$).  These two forms of feedback are the most simple physically realistic options: the potential $V_j(t)$ simply shifts in proportion to the error signal; however, because negative tunneling is difficult to achieve, quadratic feedback is the most simple realistic feedback to the tunneling term.
Both of these can be realized using quantum gas microscopes~\cite{Bakr2009,Sherson2010} where a digital mirror device can locally control both the intensity of the lattice light (increasing or decreasing tunneling~\cite{Islam2015}) as well as changing the intensity on the lattice sites (altering the on-site potential~\cite{Choi2016}).

\section{Two-site lattice}

The two-site Bose-Hubbard model with $N$ atoms can be recast as an $f=N/2$ angular momentum system with mappings $\hat n_1 - \hat n_0 \rightarrow 2\hat F_z$, $\hat a^\dagger_1 \hat a_0 + \hat a^\dagger_0 \hat a_1 \rightarrow 2\hat F_x$, and $\hat a^\dagger_1 \hat a_0 - \hat a^\dagger_0 \hat a_1 \rightarrow 2i\hat F_y$, that obey a dimensionless angular momentum commutation relation $[\hat F_i, \hat F_j] = i \epsilon_{ijk} \hat F_k$. This gives the Hamiltonian
\begin{align}
\hat H_{\mathcal S} &= -2 J \hat F_x + \Delta \hat F_z + U \hat F_z^2,\label{eq:SpinHamiltonian}
\end{align}
where we have absorbed constant terms into an overall shift in the zero of energy and defined $\Delta = V_1-V_0$.  This also results in the single filter equation 
\begin{align}
\label{eq:filter}
\tau \dot\epsilon(t) + \epsilon(t) &= 2 f_z(t)
\end{align}
for $\epsilon(t) \equiv \epsilon_1(t) - \epsilon_0(t)$ and feedback equations
\begin{align}
\Delta(t) &= \Delta_0 + G_V \epsilon(t) & {\rm and} && J(t) &= J_0 + G_J \epsilon(t)^2,\label{eq:FeedbackSpin}
\end{align}
which are slightly adjusted with an additional factor of $2$ on the right-hand side of the filter equation, resulting from the details of the angular momentum mapping.

The on-site measurement outcomes at time $t$ can be re-expressed as
\begin{align}
f_z(t) &= \langle \hat F_z(t) \rangle + m / (\sqrt{2} \varphi), \label{eq:SpinMeasurement}
\end{align}
with contribution~\footnote{The factor of $\sqrt{2}$ derives from combining two measurements of density into a single measurement of ${\hat F}_z$.} from the instantaneous expectation value $\langle \hat F_z(t) \rangle$.  
As before, the noise  is defined by a classical random variable $m$ with zero average $\overline{m} = 0$, and variance $\overline{ m^2 } = 1/2$.  Back-action is described by the Kraus operator
\begin{align}
\hat K(f_z) &=  \exp\left[-\varphi^2\left(\hat F_z - f_z \right)^2\right].
\end{align}
Taken together, these expressions provide an equivalent formulation of the two-site system in the language of angular momentum.

{\it Classical limit.} The Heisenberg equations of motion in the large-$N$ limit (i.e., ignoring quantum fluctuations and thus measurement uncertainty) reduce to the classical equations of motion
\begin{align}
\dot {\bf F} &= \left(\begin{array}{ccc}0 & -\left(\Delta + 2 U F_z\right) & 0 \\ (\Delta + 2 U F_z) & 0 & 2 J \\ 0 & -2J & 0\end{array}\right)\cdot {\bf F}. \label{eq:SpinDynamics}
\end{align}
where ${\bf F} \approx (\langle\hat F_x\rangle, \langle\hat F_y\rangle, \langle\hat F_z\rangle) $ denotes the classical angular momentum vector, ignoring quantum fluctuations. Note that because the interaction $U$ enters via a quartic term in the Hamiltonian, it enters the classical equations of motion nonlinearly. Thus in order to properly compare systems with different $f$, we keep $U f$ fixed to a constant value. Similarly, since the potential and tunneling feedback terms behave like effective fourth- and sixth-order terms in the Hamiltonian, respectively, we must also keep $f G_V$ and $f^2 G_J$ fixed to constant values.

\begin{figure}[t!]
\includegraphics{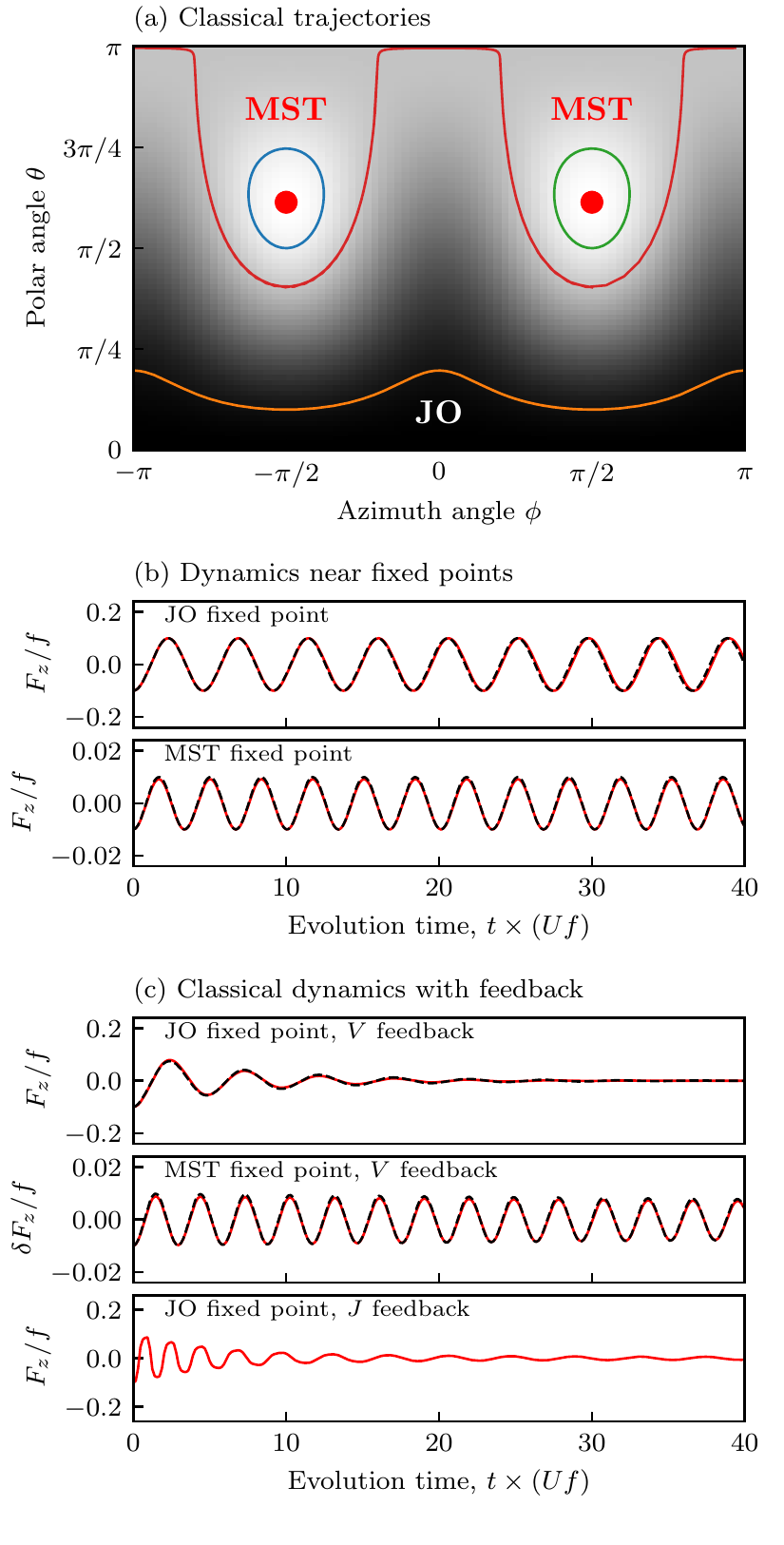}
\caption{
Classical spin model computed for $f=32$, $U f=2$, $J_0/(U f) = 0.35$, and $\Delta = 0$. 
(a) Trajectories as a function of $\theta, \phi$
with respect to the $\ex$ axis.   
These orbits are separated by the red curve into regions of Josephson oscillations (e.g., orange curve) and oscillations around the macroscopically self trapping point (e.g., blue and green curves).
The color indicates energy with black denoting the lowest energy and white the largest.
(b-c) Orbits around both JO and MST fixed points.  
Red curves denotes the result of our numerical simulation while the black dashed curves plot the result of linear response theory. 
Panel (b) plots the case with no measurement or feedback and panel (c) plots the case with feedback with $\tau=1/2$ and $t_m = 0.01$.
Feedback in $V$ [bottom two plots in (c)] damps to the JO and MST fixed points for $f G_V = -1$ and $1/4$, respectively. 
Feedback in $J$ [bottom plot in (c)] damps to the JO fixed point with $G_J = 2$.
Because the feedback is nonlinear, no linear response theory is displayed in the bottom plot.
}
\label{fig:twosite}
\end{figure}

Tunneling causes the macroscopic spin vector to precess around $\ex$; a potential imbalance leads to precession around $\ez$; and the non-linear interaction term effectively drives precession around $\ez$ with angular frequency in proportion to $F_z$.  Figure \ref{fig:twosite}(a) plots the energy [from Eq.~\eqref{eq:SpinHamiltonian}] and trajectories [from Eq.~\eqref{eq:SpinDynamics}] for $\Delta = 0$, as a function of the polar angles $\phi$ and $\theta$ with respect to an ${\bf e}_x$ aligned spherical coordinate system ($\phi=0$ in the $\mathbf{e}_y$-direction).  The displayed trajectories illustrate the two classes of orbits. (1) Near $F_x = f$ (i.e.,  $\theta=0$), the phase $\phi$ exhibits running solutions~\cite{Levy2007,Gati2007} orbiting around ${\bf e}_x$: Josephson oscillations (JO). (2) For larger $\theta$, closed orbits around $F_x = -J/U$, $F_y = 0$ and $F_z^2 = f^2-F_x^2$ correspond to oscillations about a density imbalanced excited state~\cite{Albiez2005}: macroscopic self trapping (MST), present only for $|J/U| < f$.  For $|J/U| > f$ the two MST solutions merge to form a second stable JO fixed point at $F_x=-f$.
 
{\it Linearized dynamics.} The impact of feedback is most clearly understood by first linearizing the dynamics around each of the stable fixed points of the JO and MST orbits, in general giving elliptical orbits.  The displacements $\delta F_{x,y,z}$ each obey second-order differential equations such as 
\begin{align}
\ddot \delta F_z + \omega^2 \delta F_z &= 0
\end{align}
describing motion in an effective harmonic potential, with angular frequencies
\begin{align*}
\omega^2_{\rm JO} &= 4 J_0 (J_0 \pm U f) & {\rm and} && \omega^2_{\rm MST}  &= 4 \left(U^2 f^2-J^2\right),
\end{align*}
for the JO and MST points respectively.  For $\tau=0$, the feedback described by Eq.~\eqref{eq:FeedbackSpin} becomes proportional to $F_z$ and shifts the location of the fixed points (potentially even eliminating them entirely), as well as altering the frequency and ellipticity of the orbits.  Still, the form of the differential equations is unchanged: no damping.  In the above example, conventional damping would arise from an additional friction term $\dot {\delta F_z} / \tau$, and more generally damping (or anti-damping)  will occur only when odd and even derivatives are mixed.  In our case, $\epsilon(t)$ mixes these derivatives when $\tau \neq 0$, changing the relationship between $\epsilon(t)$ and $f_z(t)$ according to Eq.~(\ref{eq:filter}), thus changing the effect of $\delta F_z$ on Eqs.~(\ref{eq:FeedbackSpin},\ref{eq:SpinDynamics}).
The effect of feedback can be quantified in linear response about both the JO and the MST fixed points, allowing us to derive a damping rate $\Gamma$ in each case, in the limit of weak damping $\omega\Gamma\ll 1$, where $\omega$ is the angular frequency of the relevant fixed point.  In this limit, the damping rates are $\omega\Gamma = \gamma \omega\tau / (1+\tau^2\omega^2)$, with the most effective damping when $\omega\tau = 1$, and where the strength $\gamma$ governs the system dependence of the feedback.  For feedback in the potential, the resulting strengths are
\begin{align*}
\gamma_{\rm JO} &= -2 J_0 f G_V & {\rm and} && \gamma_{\rm MST}  &= 2 \frac{J_0^2 G_V}{U+G_V},
\end{align*}
showing that, depending on the sign of $G_V$, the system can be damped effectively.  Focusing on the case of positive interactions $U>0$ and positive base tunneling $J_0 >0$, either only the JO fixed point is stable ($0 < -G_V < U$), only the MST fixed points are stable ($0 < G_V <U$), or both types of fixed points are stable ($-G_V > U$).

In contrast, for feedback in the tunneling channel, the quadratic response $\delta J=G_J \epsilon(t)^2$ leads to very slow damping as the density-balanced JO fixed point is approached, i.e., at the JO fixed point, $ \epsilon(t) = 0$, and the linear contribution to $\delta J$ is zero.  Thus, only the MST state has damping or anti-damping in linear response, and for the physical sign of the gain coefficient $G_J>0$, this results in anti-damping. 

{\it Fluctuations.}  Although the classical limit omits quantum fluctuations (by assuming a spin-coherent state), it need not omit measurement back action.   A prescription similar to that of Ref.~\cite{Hurst2019} for coherent states gives a final state conditioned on the random variable $m$
\begin{align}
F_{z|m} &= F_z + \sqrt{2} f \varphi m \sin^2\theta_z \label{eq:SpinUpdate}
\end{align}
for measurements of $\hat F_z$ described by Eq.~\eqref{eq:SpinMeasurement}, where $\theta_z$ is defined with respect to $\mathbf{e}_z$.  
This prescription takes an initial spin coherent state, then applies the Kraus operator; because the resulting state is not necessarily a spin-coherent state, we find the coherent state with the largest overlap as the updated state.
The core message of this expression is that $F_z$ is updated as informed by the measurement outcome, but for initial states nearly polarized along ${\bf e}_z$, the state is nearly unaltered. Physically, this is expected because we are working at fixed total angular momentum $f$, and a state already polarized along ${\bf e}_z$ cannot become more polarized.

Lastly, equating Eqs.~\eqref{eq:SpinMeasurement} and \eqref{eq:SpinUpdate} suggests an optimal measurement strength for which the true value of $F_z$ following the measurement [Eq.~\eqref{eq:SpinMeasurement}] is equal to the measurement outcome [Eq.~\eqref{eq:SpinUpdate}], i.e., the measurement outcome accurately reflects the new state of the system.  
This occurs only for an optimal measurement strength $\varphi_{\rm opt}^2 = (2 f \sin^2 \theta_z)^{-1}$, that would be selected to yield optimal performance near the desired fixed point.
We also note that with Eqs.~\eqref{eq:SpinMeasurement} and \eqref{eq:SpinUpdate}, this implies that for the optimal measurement strength, the measurement noise and back-action both scale like $f^{-1/2}$, fractionally going to zero in the $f\rightarrow\infty$ limit as one would expect.

 The low-pass filter from which we derive the error signal has time-constant $\tau$, which plays the role of an effective measurement time. 
Assuming no delay between measurements, this implies an individual-measurement coupling $\varphi_{\rm opt, 0}^2 \simeq (2 f \sin^2 \theta_z)^{-1} \times (t_m / \tau) $, so that the $\tau / t_m$ measurements which take place in the time interval $\tau$ give a single average  outcome with the optimal strength $\varphi_{\rm opt}$.

\begin{figure}[t!]
\includegraphics{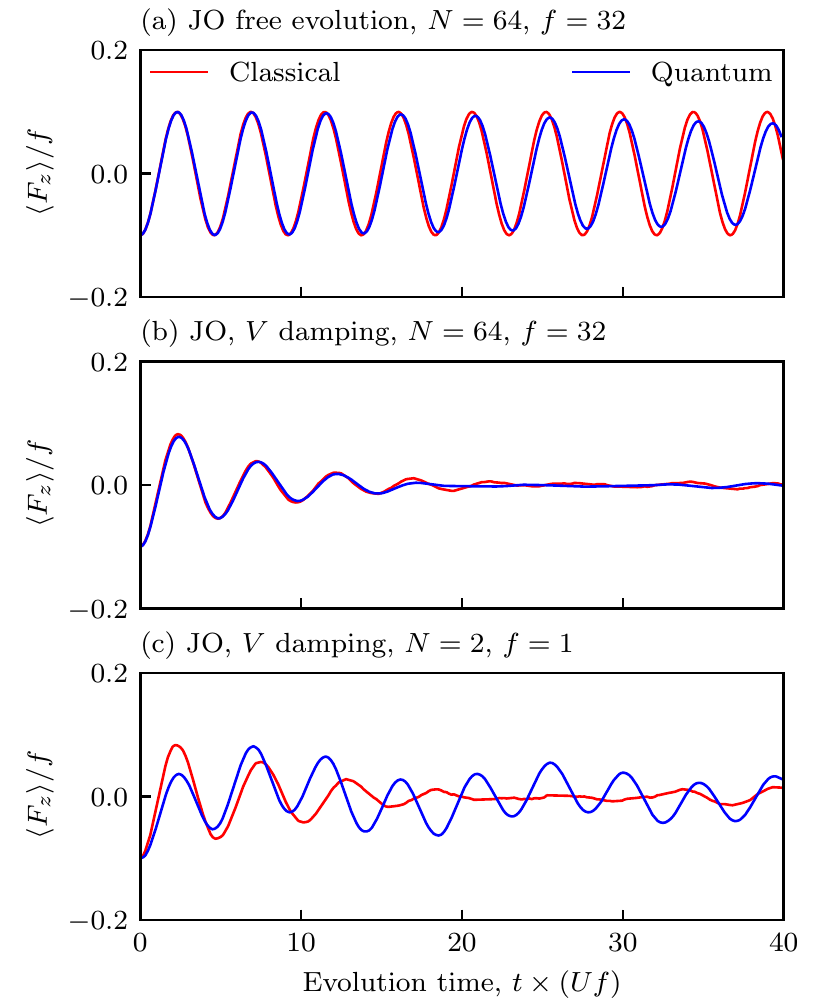}
\caption{
Comparison of classical (red) and quantum (blue) simulations with identical parameters. (a) Free evolution around the JO fixed point for the same parameters as in Fig.~\ref{fig:twosite}, including $\tau=0.5$.  (b) Evolution including measurement noise, back-action and feedback for the same parameters as in Fig.~\ref{fig:twosite}, as well as a measurement strength $f^{1/2} \varphi = t_m^{1/2}$, with $t_m = 0.01$. These trajectories were averaged over 1024 realizations to show their average behavior.  (c) Evolution for $N=2$, with parameters suitably scaled by $f$ to yield same classical dynamics as (b).
}
\label{fig:classicalquantum}
\end{figure}

{\it Quantum double well.}  With these basic insights from the classical model in mind, we compare to a quantum-trajectories stochastic Schr\"{o}dinger equation simulation of this double-well problem, including all effects of back-action resulting from measurement to the classical spin model, including measurement noise and back-action.  Figure~\ref{fig:classicalquantum}(a) shows the coherent evolution of a spin coherent state with $N=64$ or $f=32$, according to the classical spin model (red) and the quantum simulation (blue), with nearly perfect overlap.  The agreement continues to improve with increasing $N$, as neglecting quantum fluctuations becomes an increasingly good approximation.  This overlap persists in Fig.~\ref{fig:classicalquantum}(b) which includes the effects of measurement, back-action, and feedback, again showing good correspondence between the classical and quantum models.  Lastly, as the number $N$ is reduced from $64$ to $2$ (and $U$ is correspondingly increased by a factor of $64/2$ to keep $Uf = 2$), neglecting fluctuations becomes a very poor approximation, and the classical and quantum trajectories deviate.

The classical-limit modeling provided valuable insight in guiding parameter selection even for quantum systems: (1) feedback can change the dynamical steady state by introducing effective potentials; and (2) the optimal filter time $\tau$ is inversely proportional to the dynamical time scale of excitations in the measurement channel.
A basic intuition for the latter point comes from a continuously monitored classical harmonic oscillator.
The measured position converts into knowledge of momentum one-quarter period later, allowing a time-delayed feedback force to reduce that known velocity.
In the present case, $\tau$ in the lowpass filter gives the same outcome by introducing a phase shift for signals with angular frequency $\gtrsim 1/\tau$.
We certainly expect that more complex filters, proportional-integral-differential (PID) or Kalman~\cite{Kalman1960} filters for example, would give improved performance.

\section{Extended lattice}

In this section, we extend our analysis to the case of small one-dimensional Bose-Hubbard chains. Unlike the case of two sites, there is no angular momentum map, and the appropriate mean-field description is a discrete Gross-Pitaevskii equation, similar to that discussed in Refs.~\cite{Hurst2019,Hurst2020}.  Here, we focus on the quantum region by considering states with mean densities of just a few particles per-site and studying the system's behavior numerically using a quantum trajectories approach \cite{Dalibard1992,Plenio1998,Dum1992}. Correspondingly, the fluctuations in all figures are a result of sampling error.
We will also consider a range of target density distributions, in which the error signal derives from the difference between the observation and the target. 
Throughout, we will express everything in units where the interaction strength $U=1$.
Recent closely-related work investigates preparing similar Mott-like states via global measurements rather than the local measurements considered here~\cite{Wu2021}.

\emph{Feedback models.}
We consider two possible tunneling feedback schemes. The first is the nearest-neighbor approach introduced in Eq.~\eqref{eq:FeedbackJ}. The second, the ``imbalance approach'', uses highly non-local information, and relies instead on imbalances between the left and right sides of the system. For the double-well system considered earlier, these two schemes are identical. The tunneling strengths are, respectively, 

\begin{equation}
J_j(t) = G \big|[\epsilon_j(t)-\epsilon_{j+1}(t)]-[N_j-N_{j+1}]\big|^2,
\end{equation}
\begin{equation}
\label{JNN}
J_j(t) = G \big|[\epsilon_{j,L}(t)-\epsilon_{j,R}(t)]-[N_{j,L}-N_{j,R}]\big|^2,
\end{equation}
where
\begin{equation}
\begin{aligned}
N_{j,L} = \sum_{k \leq j} N_k, && N_{j,R} = \sum_{k > j} N_k, \\
\epsilon_{j,L} = \sum_{k \leq j} \epsilon_k, && \epsilon_{j,R} = \sum_{k > j} \epsilon_k,
\end{aligned}
\end{equation}
and $N_j$ indicates the target occupation number of site $j$. Since the tunneling strength without feedback is $J_0 = 0$ and measurement occurs in the number basis, this means that the target state $|N_1 N_2 \cdots \rangle$ will be an evolution-free fixed point in the absence of feedback induced by noise from the measurements. Given the structure of the imbalance approach, which entails separating the system into left and right sides, we will consider open boundary conditions when discussing state preparation.

The purpose of the imbalance approach is to take advantage of the global knowledge of the target state, knowledge that is not used in the nearest-neighbor approach. For example, consider the scenario in which the target state is $|4 3 2 1 0 \rangle$ and the system is in the state $|0 4 3 2 1 \rangle$. In this case, the nearest-neighbor approach will only initially turn on the tunneling between the first two sites and gradually turn on the other tunneling terms from left to right as the bosons move to the left. In contrast, the imbalance approach will turn on all tunneling terms and begin to transfer atoms from the right side of the system to the left. However, the use of this more global knowledge comes at a cost to the feedback uncertainty. Since $\epsilon_{j,L/R}$ involves a sum of $N$ different measurement records, the resulting uncertainty will be enhanced by a factor of $\sqrt{N}$. As a result, there is a trade-off between using global knowledge and the uncertainty in the feedback. Finally, we will discuss these two feedback approaches in terms of a continuous measurement rate $\kappa$, where $\varphi \equiv \sqrt{2 \kappa t_m}$ and we use $t_m = 0.01$ in our numerical simulations. 

\emph{State preparation.} 
We compare the performance of these two approaches for three different target states: $\psi_1 = |11111\rangle, \psi_2 = |20202\rangle$, and $\psi_3 = |30003\rangle$. We fix the filter time $\tau = 0.3$ and the evolution time $T = 10 + \tau$ \footnote{The extra time $\tau$ is to allow for an initialization of $\epsilon_j$ before the feedback starts being applied at $t = \tau$. 
} in units of $U^{-1}$ and examine the behavior of the final target state overlap $|\langle \psi_i| \psi(T) \rangle|^2$ as a function of $\kappa$ and $G$. For each trajectory, we initialize the system in a Haar random state. While we could consider the steady-state behavior of this system, from the perspective of creating a target state in a physical system, it is more useful to consider the finite-time behavior.
\begin{figure}[t]
\includegraphics{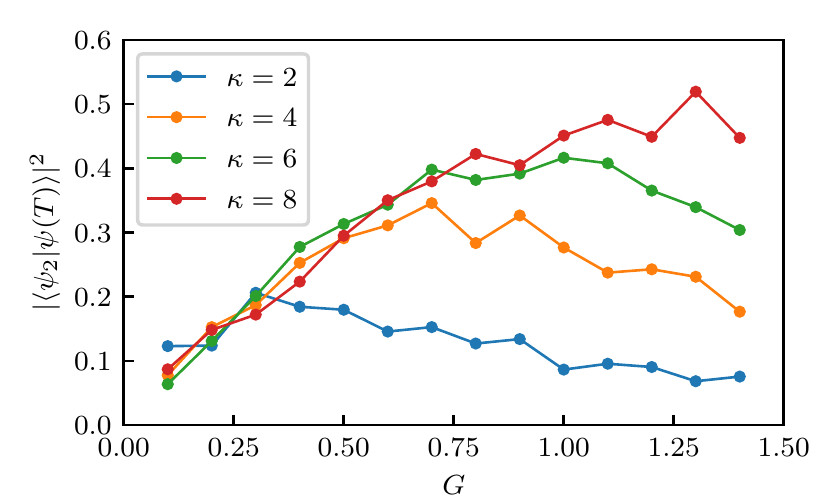}
\caption{Average overlap $|\langle \psi_2| \psi(T) \rangle|^2$ for different values of $G$ and $\kappa$ for the nearest-neighbor approach using 256 trajectories. \label{20202Fig}}
\end{figure}

\begin{figure}[t]
\includegraphics{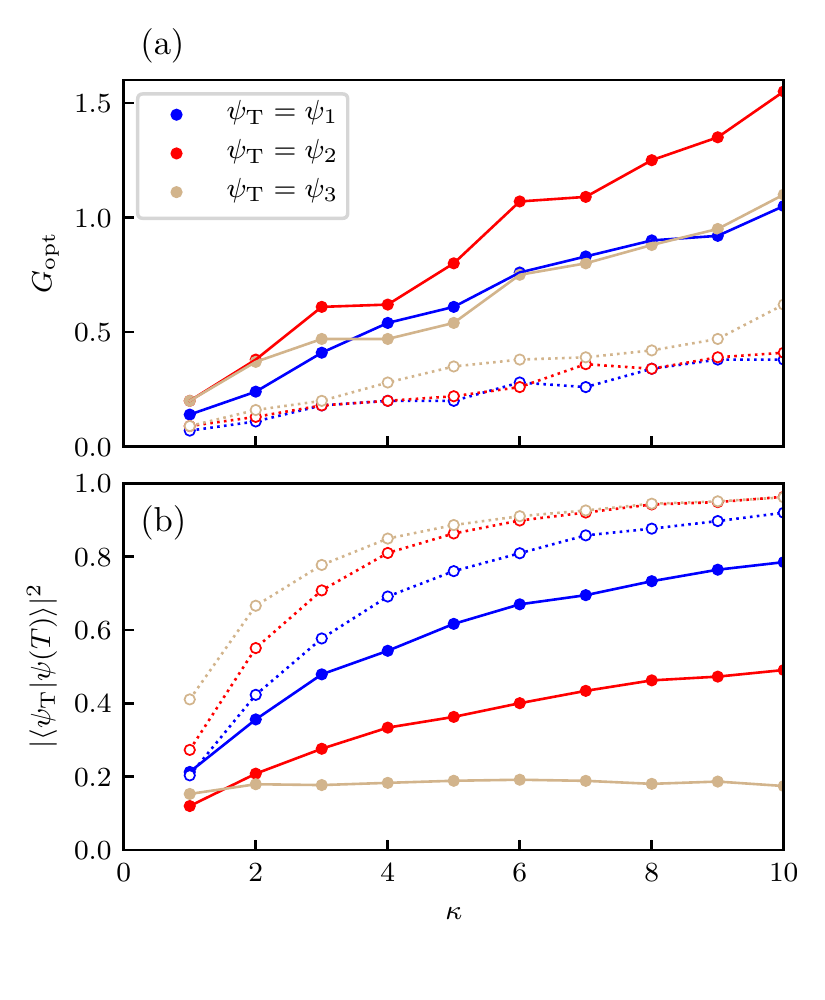}
\caption{Optimal behavior of the two feedback approaches using 2048 trajectories for each data point except for the nearest-neighbor approach for $\psi_3$, which uses 4096 trajectories. (a) Optimal gain $G_{\rm opt}$ as a function of $\kappa$. (b) Optimal target state overlap as a function of $\kappa$. The solid (dotted) lines correspond to the nearest-neighbor (imbalance) approach. Blue, red, and tan lines correspond to $\psi_T = \psi_1, \psi_2, \psi_3$, respectively.}
\label{optimalFig}
\end{figure}

Fig.~\ref{20202Fig} illustrates the target state overlap for $\psi_2$ at different values of $\kappa$ and $G$ using the nearest-neighbor approach. The behavior for other target states and the imbalance approach are qualitatively similar. For all values of the measurement strength $\kappa$, we see that there is an optimal choice of the gain $G$ for which the overlap is maximized. This optimal value of the gain represents a balance between the need to direct the system to the desired target state quickly and the fact that measurement uncertainty can lead to ``errors'' in the application of the feedback. We identify the optimal value of the gain and the corresponding target state overlap for the different target states and measurement strengths in Fig.~\ref{optimalFig}. Several trends can be identified in the optimal feedback behavior. 

The first trend is that the optimal gain grows linearly with the measurement strength. This reflects the fact that stronger measurements correspond to more accurate knowledge of the system's state, so the system can be more strongly driven to the target state without making errors due to the measurement uncertainty. Additionally, the gain must increase with the measurement strength to avoid quantum Zeno effects, i.e., the dynamics due to the feedback must become sufficiently fast so that the repeated weak measurements do not effectively become a strong measurement before the system can evolve coherently. 

The second trend is that optimal gain is larger for the nearest-neighbor approach than the imbalance approach. This is most likely a result of the fact that the uncertainty is larger for the imbalance approach. Since the imbalance approach uses 5 measurements for each tunneling link  while the nearest-neighbor approach uses 2, the relative uncertainties in the tunneling strength differ by roughly a factor of 2.5 \footnote{for white noise $\xi$ with standard deviation $\sigma$, the mean and standard deviation of $\xi^2$ are $\sigma^2$ and $\sqrt{2} \sigma^2$, respectively, hence the factor of $2.5$ rather than $\sqrt{2.5}$}, which is consistent with the observed behavior. 

The third trend is that for the parameters and target states considered, the imbalance approach performs much better than the nearest-neighbor approach. The nearest-neighbor approach performs better the more homogeneous the target state is. However, this begins to change at lower values of the measurement strength, where the target state overlap sharply begins to fall. This is a consequence of the fact that, at sufficiently small measurement strengths, the uncertainty in the measurement---and therefore the feedback signal---overwhelms the useful information. Since the imbalance approach involves more measurements, this behavior occurs earlier than for the nearest-neighbor approach. 

\emph{Nearby state preparation.} 
For larger systems, the effect of the increased uncertainty that comes with the imbalance approach becomes far more important. A natural approach to avoid this issue is to use a more hierarchical approach to the application of feedback. When the system starts initially far from the desired target state, a more global approach to feedback should be used. As the system gets closer to the target state according to the measurement record, then increasingly local feedback approaches can be used. 

\begin{figure}[t!]
\includegraphics{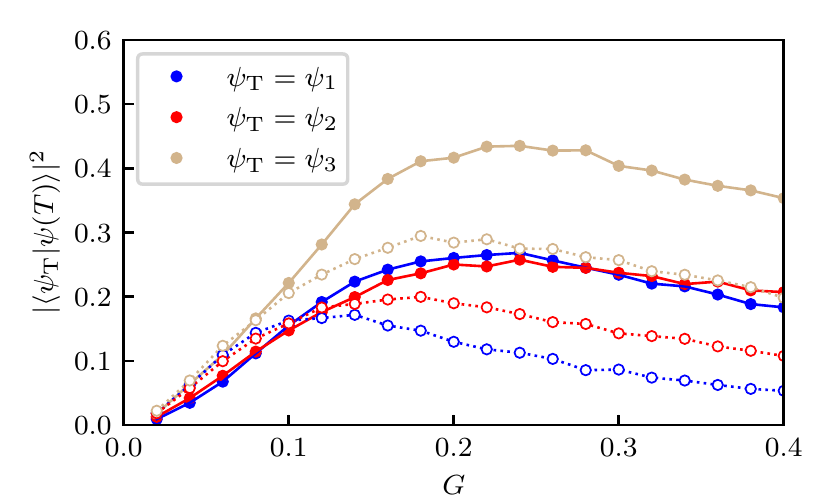}
\caption{Average overlap $|\langle \psi_{\rm T}| \psi(T) \rangle|^2$ for different values of $G$ when the initial states are close to the target state using 4096 trajectories. The solid (dotted) lines correspond to the nearest-neighbor (imbalance) approach. Blue, red, and tan lines correspond to $\psi_T = \psi_1, \psi_2, \psi_3$, respectively. \label{ShortFig}}
\end{figure}

While the systems we can consider numerically are too small to apply such a hierarchical feedback approach, we can explore how the performance of the two feedback approaches can change if the system is already closer to the target state. To do so, we consider initial states whose only difference from the target state is that a single boson has been moved one site away. Additionally, we allow the feedback to be applied for only one unit of time after the initialization of the measurement record (which requires time $\tau$), reflecting the fact that the local feedback would be applied for a shorter time in this hierarchical approach. The results of our numerics are shown in Fig.~\ref{ShortFig} for $\kappa = U$, demonstrating that the nearest-neighbor approach performs better when the system is already close to the desired state due to the reduced uncertainty in the feedback. Note that this is in spite of the fact that the nearest-neighbor approach will initially turn on three coupling terms while the imbalance approach will turn on only one in the absence of measurement uncertainty.

\begin{figure}
\centering
\includegraphics{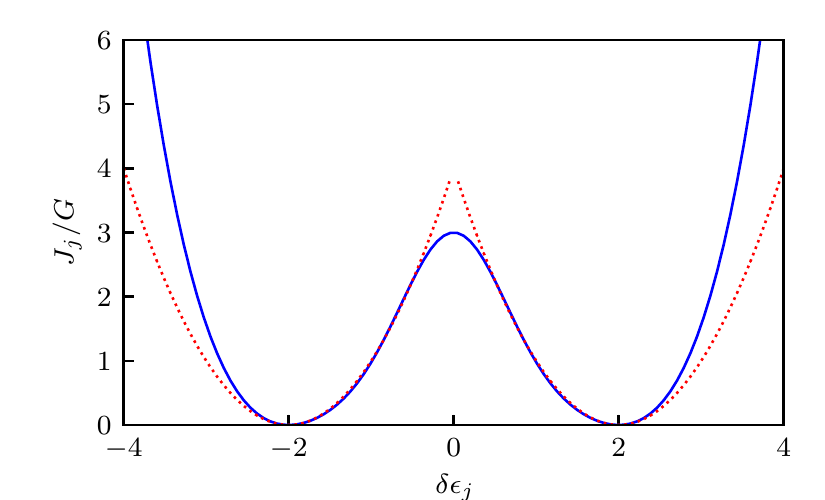}
\caption{Solid blue line shows tunneling feedback used for symmetry breaking transition as a function of $\delta \epsilon_j$. Red dotted lines correspond to the feedback used in the state preparation section, with the left (right) curve corresponding to the target state $|020202\rangle$ ($|202020\rangle$).} \label{symbreaktunnelfig}
\end{figure}

\emph{Symmetry-breaking transition.} 
The emergence of phase transitions via feedback has been investigated in a variety of diverse contexts~\cite{Grimsmo2014,Kopylov2015,Mazzucchi2016,MunozArias2020b, Ivanov2020, Hurst2020,Kroeger2020,Ivanov2020b,Ivanov2021}, and here we explore a symmetry-breaking transition in an extended Hubbard chain.
To observe this, we will consider a six-site lattice with periodic boundaries whose tunneling feedback drives the system towards the states $|202020 \rangle$ or $|020202 \rangle$. Due to the periodic boundary conditions and type of symmetry breaking, the left-right imbalance approach is not applicable, so we consider a generalization of the nearest-neighbor feedback. In particular, the tunneling feedback is 
\begin{subequations}
\label{symbreaktunnel}
\begin{equation}
J_j(t) = \frac{G}{16} \left(1 + 2 e^{-\delta \epsilon_j^2}\right)(\delta \epsilon_j-2)^2(\delta \epsilon_j+2)^2,
\end{equation}
\begin{equation}
\delta \epsilon_j = [\epsilon_{j}(t)-\epsilon_{j+1}(t)],
\end{equation}
\end{subequations}
with $\epsilon_{7} \equiv \epsilon_1$. This is essentially a product of the tunneling feedback in Eq.~(\ref{JNN}) for $N_j - N_{j+1} = \pm 2$. The Gaussian term is included in order to increase the tunneling for $\epsilon_{j}-\epsilon_{j+1}\approx 0$ compared to just the quartic potential so that it is comparable to the corresponding tunneling when only one of the two aforementioned states is a target state. This will serve as a barrier which hinders the system from moving from $|202020\rangle$ to $|020202 \rangle$ and vice-versa (see Fig.~\ref{symbreaktunnelfig}).

An additional important feature to note for this choice of feedback is that these are not the only two states which will lead to small tunneling. States of the form $|0 2 4 2 0 0\rangle$ can also be relatively stable. While there is one pair of sites with large tunneling, this pair of sites has no bosons, so the large hopping is not sufficient on its own to change the state. However, there are two factors that can make such states unfavorable. The first is that the creation of these states requires four bosons on a single site, which is energetically unfavorable. The second is that once measurement-induced fluctuations cause one of the bosons at sites with two bosons to move into an empty site, the boson will quickly move to the next site. 

In order to investigate the symmetry-breaking behavior of this feedback, we consider the behavior of the steady-state density matrix $\rho_{ss}$ for fixed $\kappa$ as a function of $G$, which we obtain via a combination of ensemble and ergodic averaging. Additionally, we define an effective Hamiltonian $H_{ss}$ according to $e^{-H_{ss}} \equiv \rho_{ss}$. In analogy to equilibrium systems, a symmetry breaking phase will occur when the two lowest energy levels of $H_{ss}$ become gapped from the rest of the spectrum. In terms of $\rho_{ss}$, the two corresponding states will have a much larger probability than the rest of the eigenstates, and the gap in $H_{ss}$ describes the exponential suppression in probability of these other eigenstates. Hence in the absence of a gap, no particular configuration is favored, and the system appears disordered.

\begin{figure}
\centering
\includegraphics{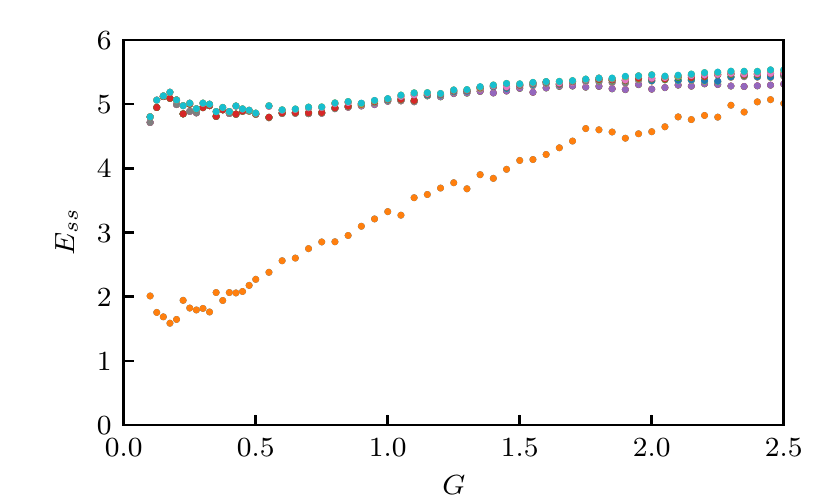}
\caption{Twenty smallest effective energies of steady-state $H_{ss}$ as a function of $G$. Eigenvalues are calculated after symmetrizing $\rho_{ss}$ with respect to translations and reflections, leading to degeneracy for states related through these symmetries; quantitatively similar behavior emerges without symmetrization, although the eigenvalues are no longer as degenerate. The orange dots are doubly degenerate and correspond approximately to $|202020\rangle, |020202\rangle$, with the strongest overlap occurring for the larger gaps. For $G < .5 U$, $\rho_{ss}$ is averaged over 128 trajectories from $t=100 U^{-1}$ to $t = 400 U^{-1}$. For $G \geq .5 U$, $\rho_{ss}$ is averaged over 32 trajectories from $t = 20 U^{-1}$ to $t = 320 U^{-1}$. The initial time delay for averaging ensures that the system has relaxed to the steady state.\label{ESteady}}
\end{figure}

Fig.~\ref{ESteady} plots the eigenvalues of $H_{ss}$ as a function of $G$ for $\kappa = 5 U$ using the initial state $\psi(0) = |111111\rangle$, which will prefer both symmetry-breaking states equally, although we expect the same steady state for Haar random initial states. We see that an effective gap opens for a range of $G$, with two effective energies much smaller than the rest. As expected, these two eigenvectors correspond approximately to $|202020\rangle$ and $|020202\rangle$, while the other possible stable states (e.g., $|024200\rangle$) are not strongly occupied. Additionally, as this gap increases, the auto-correlation times increase, necessitating more sampling for larger gaps. This corresponds to the fact that when the system is in either of the two symmetry-broken states, the feedback makes it difficult to leave this state, so the system spends a long time in the same state, thus slowing down the ergodic sampling of $\rho_{ss}$. In the limit of large $G$, even small fluctuations will drastically modify $J_j(t)$, preventing the system from preferring any particular state and causing the gap to close.

As $G$ is increased, this gap eventually closes and does not reopen. Physically, this is because in the limit of large $G$, the hopping becomes very large, so small fluctuations in the measurements lead to large fluctuations in the hopping, preventing the system from preferring any given state. This is qualitatively similar to an infinite temperature limit in which no particular state is energetically preferred. In the limit of small $G$, the gap similarly closes because the measurement strength overwhelms the feedback, causing the system to relax to a random number state. The start of this closure can be observed in Fig.~\ref{ESteady}, although due to the quantum Zeno effect, identifying the steady state for smaller values of $G$ becomes numerically prohibitive as it requires increasingly long relaxation times. 

\section{Outlook}

This study combines concepts from many-body physics with the techniques of quantum control and feedback to create and characterize low entropy many-body dynamical steady states.  Creating such dynamical steady state in the laboratory environment hinges on exploring the implication of experimental realities, for example: What are the consequences of imperfect detection?  Of limited spatial resolution?  What is the impact of limited feedback bandwidth?  Real implementation must face these questions head-on.

From a foundational perspective, it remains to be seen if there exist dynamical steady states which are forbidden in associated thermal equilibrium systems. 
For example, nonreciprocal couplings can be realized using feedback, which can lead to rich non-equilibrium phenomena \cite{Metelmann2015,Metelmann2017,Young2020,Fruchart2021}. Alternatively, feedback that is non-local in time can be used to engineer non-Markovian baths even though the measurements themselves are Markovian. Similarly, even with local measurements of density, non-local feedback can emulate aspects of long-range potentials \cite{Muschik2013}.
Stochastic descriptions such as ours can be reframed in terms of the so-called feedback master equation~\cite{Wiseman2009}, in which the time delay introduced by the low pass filter introduces non-Markovian terms that cannot be described in a Lindblad form.
This at least provides the opportunity for creating dynamics and steady states that are not achievable using realistic Lindblad terms. 
Moreover, even if the steady state may look thermal, its linear response may still violate the fluctuation-dissipation theorem \cite{Callen1951,Kubo1966}, indicating the persistence of non-equilibrium features. Additionally, new types of dynamics are possible when the bosons include a spin degree of freedom, such as quantum Zeno-like effects which can emerge if one spin state is subjected to stronger measurements than the other.

While we have demonstrated how the use of non-local feedback can be employed to prepare desired target states more efficiently than local feedback, this was done for the relatively simple case of Mott-like number states. An interesting next direction, then, is to consider preparing more complex states, such as superfluid-like states with long-range coherences. By adding density modulations through the use of non-local feedback as we did for the number states, this opens the possibility of preparing supersolid and supersolid-like states \cite{Chester1970,Boninsegni2012,Leonard2017,Li2017a}. More broadly, this would provide further insight into how the use of measurement and feedback affects the ability to realize intrinsically quantum dynamics.

Although we have illustrated how a symmetry-breaking phase transition may in principal emerge due to measurement and feedback in a many-body quantum system, several open questions remain. Here, our analysis has been restricted to small 1D chains, so an important question is whether this corresponds to a phase transition in the thermodynamic limit. Even if it does not exist in 1D, an analogous transition may exist in higher-dimensional systems. Additionally, understanding the behavior of critical phenomena in these systems is another important direction, such as whether they are equivalent to a quantum phase transition, a classical phase transition, or something entirely different. In Ref.~\cite{Ivanov2020}, it was shown that in zero-dimensional systems, modifying the form of non-Markovian feedback can lead to changes in the critical exponents of the phase transition, so similar phenomena may arise in a many-body context as well. 

A natural direction to explore in order to investigate these questions are measurement-induced phase transitions, which have been the subject of intense theoretical research in recent years and are defined by the scaling behavior of the entanglement entropy \cite{Li2018,Li2019,Skinner2019,Chan2019,Choi2020,Jian2020,Turkeshi2020,Gullans2020,Ippoliti2021,Lavasani2021,Nahum2021,Minato2021,Cao2019,Szyniszewski2019,Bao2020,Goto2020,Alberton2021}, with recent initial experimental evidence \cite{Noel2021}. Aside from the key difference that these systems are not subject to feedback, these often involve coherent evolution defined by random unitary circuits and strong measurements, although similar behavior has been shown to emerge for weak measurements as well \cite{Cao2019,Szyniszewski2019,Bao2020,Goto2020,Alberton2021}. A natural question is how these phase transitions can be modified through different applications of feedback; will the phase transition only shift, or can qualitatively new physics emerge?

Another promising approach to exploring the above questions is to investigate the similarities that systems subject to measurement and feedback have with driven-dissipative systems, which are systems that are subjected to coherent drive and incoherent dissipation that have been studied extensively in a variety of contexts \cite{Diehl2008,Szymanska2007,Yi2012,Bardyn2013, Baumann2010,Carr2013a,Malossi2014,Rodriguez2017,Fitzpatrick2017,Dogra2019,Jin2016,Minganti2018,Soriente2018,Torre2010,Gopalakrishnan2010, Tauber2014a,Altman2015,Sieberer2016, Torre2013,Maghrebi2016,FossFeig2017, Marino2016,Paz2021,Rota2019,Iemini2018, Nagy2019,Young2020,Roberts2020,Scarlatella2021,Sierant2021}. From an abstract standpoint, these two types of systems are very similar, with dissipation playing a role analogous to continuous measurements and drive playing a role analogous to the feedback. As a result, insights or techniques from studying one type of system can lead to insights in the other. For example, phase transitions in driven-dissipative systems have been studied extensively using a Keldysh-Schwinger and functional integral formalism \cite{Torre2010,Gopalakrishnan2010, Tauber2014a,Altman2015,Sieberer2016, Torre2013,Maghrebi2016,FossFeig2017,Marino2016,Young2020, Paz2021}, so it is important to explore how these same techniques can be applied to systems subject to measurement and feedback. Similarly, effective classical equilibrium criticality is observed generically in many driven-dissipative phase transitions \cite{Torre2013,Maghrebi2016,FossFeig2017}, with some important exceptions \cite{Marino2016,Rota2019,Young2020,Paz2021}, so measurement-feedback systems may provide new avenues to realize novel forms of non-equilibrium criticality and quantum criticality. Moreover, recent research indicates that dissipative phase transitions and measurement-induced phase transitions need not coincide \cite{Sierant2021}, which means that measurement-feedback phase transitions may similarly lead to distinctive forms of criticality.

\begin{acknowledgments}
We benefited greatly from conversations with H.~M.~Hurst, J.~K.~Thompson, J.~Ye, L.~Walker, R.~Lena, A.~Daley, L.~P.~Garc\'{i}a-Pintos, S.~Guo, and E.~Altuntas.
We appreciate the careful reading from S.~Lieu and D.~Barberena. J.T.Y.~was supported by the NIST NRC Research Postdoctoral Associateship Award. A.V.G.~acknowledges funding by DARPA SAVaNT ADVENT, AFOSR, ARO MURI, AFOSR MURI, U.S. Department of Energy Award No.~DE-SC0019449, NSF PFCQC program, DoE ASCR Accelerated Research in Quantum Computing program (Award No.~DE-SC0020312), and DoE ASCR Quantum Testbed Pathfinder program (Award No.~DE-SC0019040). This work was supported by NIST. Analysis was performed in part on the NIST Raritan HPC cluster.
\end{acknowledgments}

\bibliography{MeasurementFeedback}

\end{document}